# Towards Understanding What Contributes to Forming an Opinion

Peng Wang[1,2], Jia Song[1], Jie Huo[1,3], Rui Hao[1,3] and Xu-Ming Wang[1,3,*]

1 School of Physics and Electrical Information Engineering, Ningxia University, Yinchuan, 750021, PR China

2 Department of General Studies, Beifang University of Nationalities, Yinchuan, 750021, PR China

3 Ningxia Key Laboratory of Intelligent Sensing for Desert Information, Yinchuan, 750021, PR China

* E-mail: wang_xm@126.com

## Abstract

Opinion evolution mechanism can be captured by physics modeling approaches. In this context, a kinetic equation is established by defining a generalized displace (cognitive-level), a driving force and the related generalized potential, information quantity, altitude. It has been shown that the details of opinion evolution depends the type of the driving force, self-dominated driving or environment-dominated driving. In the former case, the participants can have their altitudes changed in the process of competition between the self-driving force and environment-driving force. In the latter case, all of the participants are pulled by the environment. Some regularity behind the dynamics of opinion is also revealed, for instance, the information entropy decays with time in a special way, etc. the results may help us to get some deep understandings to formation of a public opinion.

## 1 Introduction

In the past decades, opinion evolution or formation has attracted a good deal of attention in physics communities [1-17]. An opinion often evolves coupling with a particular social structure, in which there will be a complicated interaction between different participants in aspects of knowledge, experience, custom and value orientation, etc. The interaction might lead the opinion to different effects, coincidence or conflict.

To investigate the dynamics of opinion, some models in framework of non-equilibrium statistical physics are proposed. One of the most popular is the variants of the Ising model to simulate evolution of binary opinions [1-5]. A cellular automata model, that is described as a "missionary" model, presented the dominating rules: a pair of neighbors who hold to parallel orientation may attempt to persuade their nearest neighbors to follow the same orientation and they will get their wish; while the pair who have antiparallel orientation cannot get the wishes and its two neighbors have the orientation that is contrary to that persuaded by one of the pair, respectively [2]. This model was then transformed into a two dimensional lattice model, which investigated the reasonable performing rule: all of the neighbors to neighboring pair that has a parallel orientation can take the advice, while the neighbors to the pair that has antiparallel orientation will keep changeless since each of them is forced by two "voices", and therefore is at a loss as to what to do [ 3]. The "missionary" model was also modified into the one to simulate price formation in a financial market as the spin is interpreted as the bullish altitude or bearish altitude [4]. The spirits of the aforementioned model was extended to investigate an opinion formation in a general process that surpasses a choice of either this or that between "rightist" or "leftist", denoted by the up spin and down spin [5]. The most attractive cellular automata model is that suggested by Castellano [1], which emphasizes fertilization of culture between the neighboring regions, and uncovered the non-equilibrium phase transition from an order phase marked by culture polarization to a disorder phase characterized by culture fragmentation.

To explain opinion formation in social groups, an agent was regarded as a particle surrounded by others, and the kinetic theories (kinetic equations) describing the collision between particles was explanted to simulate the interaction between the participants in the typical situations such as general choice formation [6], a ballot in an election or contest [7], opinion formation under the influence of some strong leaders [8], and so on [9-17]. It is worth noting that the interaction in Ref. [8] denotes opinion exchange or self-thinking, and steady solution of the Fokker-Plank equation that is derived from the kinetic equation presents the emergence or decline of opinion leaders.

As far as the physics mythologies used in researching opinion evolution are concerned, there are cellular automata [1-5, 7, 9, 10], random dynamics [6, 11, 12], statistical dynamics [8], mean field [1] and other methods of analogy between physical system and social system [13-17]. However, from the point of physical view, the fundamental factors that drive the evolution of opinion have not yet been revealed completely. Fortunately, the investigations conducted by Helbing [18, 19] and Martins [20, 21] might provide us with inspiration for the further studies on the subject. Based on the belief that there exists a principle, which may be similar to the Newton second law, governing the social system, the investigations try to uncover physical essence of opinion dynamics via exploring the correspondences of driving force, acceleration, mass/inertia in opinion formation process.

In this article, we try to establish a model to simulate the formation of a public opinion, and to find what essentially contributes to the consensus. That is, we suggest a kinetic equation for a participant described by driving force (from both himself/herself and environment mainly characterized by the altitudes of the participants and quantity of information delivered in the communication), displace (cognitive-level) and corresponding acceleration. The simulation presents the formation of opinion and the characteristics relating to driving type, self-dominated or environment-dominated. The variation of entropy with time is calculated to interpret the difference between these two types. The driving forces are expressed in a general form.

## 2 The Cognitive Model

Formation of a public opinion is regarded as one of the typical behaviors of a social group. The group is also taken as a complex open-system, of which the participants interact with each other to form a complex structure. However, due to the difference in culture background, education experience and age, the participants may have different reaction to an event or a thing, and learn the new conception with different cognitive style and perspective. In other words, if one has a relatively high cognitive-level, he/she should have high culture literacy and a wide field of view, which mean a strong judgment for the participant. In contrast, if one has a relatively low cognitive-level, he/she lack clear understandings to new things due to the poor judgment and therefore is short of clear understand to the nature of things. With the development of new thing and the change of living environment and education experience, the individual's cognitive-level could be improved considerably. If one's cognitive-level, x, increased, it means that the agent gets positive energy from the environment, and will enhance his/her perceptibility. In contrast, if variable, x, is reduced, it means, on the one hand, that the individual often shows decision-making-mindlessly in the face of new things ; on the other hand, individual bears a suspicion altitude or resistance to the new thing, which lead to decline of the judgment to new things. In short, different individual follows different evolution modes that depend not only on the participant's status but also its surroundings to exchange information. Thus, we establish a model for the evolution of cognitive-level, which can be regarded as a generalized displace here, for the i-th participant.

$$\ddot{x}_i = g_i(t) + f_i(t) + \xi_i(t) \tag{1}$$

Where t is time, $g_i(t)$ represents so-called self-driving force from the $i$-th participant himself, $f_i(t)$ denotes so-called environment-driving force which originates in the interactions between the participant and each of the others, and variable $\xi_i(t)$ denotes the Gaussian white noise that is characterized by mean value,

$$\langle \xi_i(t) \rangle = 0 \tag{2}$$

Now let's discuss expressions of the self-driving force and environment-driving force, respectively. A force is generally determined by the gradient of a potential, which is also generally defined by displace. How do we define the generalized potential? To answer this question, we should concentrate on the related aspect, i.e., the essence of the cognitive-level. From recognition viewpoint, the higher the cognitive-level one has, the more information one can obtain from a particular thing. Therefore, the quantity of information, for a particular person and for a particular thing, might positively correlate with the cognitive-level, so we introduce an ansatz function to describe the dependence of the quantity of information on the cognitive-level. It reads

$$h_i(x_i) = e^{Cx_i}, \tag{3}$$

where c is a constant. It may be also crucially important for defining the generalized potential that both of the driving force and the variation of the cognitive-level depend not only the quantity of information but also the altitude of the agent to the particular thing. Then the generalized potential can be defined by the product of quantity of information $h_i(x_i)$ and altitude $\sigma(t)$, and the self-driving force is expressed as

$$g_i(t) = A\nabla \big( h_i(x_i(t))\sigma_i(t) \big), \tag{4}$$

where $A$ is a strength coefficient that weighs the contribution of self-driving force to the gross driving force. It is evident that two forces must be opposite ones if the corresponding altitudes are opposite and the quantities of information are same. It follows that the definition of self-driving fore may be rational.

The environment-driving force is an aggregate of the influences from all of the participants but the $i$-th agent. In other words, each of the participants likely makes a contribution to the potential due to the interactions between the $i$-th and the others. It is similar to that presented in equation (4), the potential, corresponding to the interaction between the $i$-th participant and $j$-th one, not only depends on the quantity of information included in the thing for the $j$-th participant but also relates to the altitudes of the two parties. Nay more, the interaction only takes effect under the condition of that the information obtained by the $j$-th participant is delivered to the $i$-th. So the potential can be written as

$$U_{ij}(t) = P_{j \to i} h_j(x_j(t))\sigma_i(t)\sigma_j(t), \tag{5}$$

Where $P_{j \to i}$ denotes the probability of the information transmitted from the $j$-th participant to the $i$-th. The Fermi's distribution [22-24],

$$P_{j \to i} = \frac{1}{1 + e^{-D(x_j - x_i)}}, \tag{6}$$

is recommended here based on the fact that in the real world a person is inclined to receive the opinion of those who (authorities or specialists) have knowledge of a particular thing richer than himself/herself. That is, the more difference of cognitive-level between $j$-th participant and the $i$-th, the more effective information (may be novel for the $i$-th) is transmitted to the $i$-th, and the greater the influence exerting by the $j$-th on the $i$-th will be. So the probability $P_{j \to i}$ is negatively correlated to the difference of cognitive-level $x_j - x_i$. And we may as well adopt the form of the Fermi's distribution which can demonstrate the negative correlation. It must be emphasized that the interaction potential $U_{ij}$ is determined by the coupling of the altitudes of the two participants. The specific effect of the coupling depends on the sign of altitude $\sigma_i(t)$ and that of $\sigma_j(t)$, which are denoted by the product of the two sign function in equation (5). To the $i$-th participant, this product embodies the acceptance and rejection of the transmitted information from the $j$-th. The acceptance occurs just as the two participants have same altitude, while the rejection appears as their altitudes are opposite. So, the environment-driving force can be written as

$$f_i(t) = B \nabla \left( \frac{1}{N-1} \sum_{j \neq i, j=1}^{N} U_{ij}(t) \right), \tag{7}$$

Where $B$ is a strength coefficient that associates with environment effect on the $i$-th participant. And we introduce the factor, $\frac{1}{N-1}$, to modify the sum of interaction potentials to be a reasonable quantity that can be matched with the potential that is associated with the self-driving force based on the basic fact that there often be competition between the self-driving and the environment-driving other than an obvious asymmetry of the two forces.

In this process, the altitude of a participant may also adjust due to the influence of the other's altitudes. It can be regarded as continuously changing based on the consideration that the continuous vale of the altitude can represents the size of one's belief, and its variation range may be confined within region $[-1,1]$. So we have

$$\sigma_i(t+\tau) = \sigma_i(t) + \frac{1}{N-1} \left( \sum_{j=1, j \neq i}^{N} \frac{1}{1+e^{-D(x_j(t)-x_i(t))}} \right) \qquad \sigma_i \in [-1,1] \tag{8}$$

Where $\tau$ denotes the time internal. Formula $\frac{1}{N-1} \left( \sum_{j=1, j \neq i}^{N} \frac{1}{1+e^{-D(x_j(t)-x_i(t))}} \right)$ presents the accumulating effect of the altitude from all of the participants in the environment and factor $\frac{1}{N-1}$ is introduced based on a consideration similar to that one in equation (7). And $\mathrm{Mod.}[-1,1]$ stands for the modulo operator with value of the altitude in range $[-1,1]$. It is also mentioned above that the plus sign implies a positive altitude of the participant to a certain thing (acceptance), while the minus sign means a negative altitude (rejection).

## 3 Numerical Simulation and Discussions

Let's perform a numerical simulation by the cognitive model with a set of fixed parameters ($C = 0.16, D = 0.99$) and under initial conditions that altitudes and cognitive-levels are randomly evaluated in range $[-1,1]$ and $[0,15]$, respectively. And then the participants ($N = 900$) are initially randomly divided into two types. One has 446 members with plus altitude and the other has 454 members with minus altitude.

### 3.1 Opinion Formation and Driving Forces

To investigate, in detailed, the opinion evolution process, we focus on the influence of the self-driving force and the environment-driving force on the opinion formation and some related aspects.

In point of driving strength, there evidently exist two typical situations, i.e., the self-driving force is greater than the environment-driving force and is called self-dominated driving, and the opposite situation that is called environment-dominated driving. Fig.1 presents the results of a self-dominated driving case calculated as the strength parameters are chosen as $A = 7.5 \times 10^{-3}, B = 2.5 \times 10^{-3}$. Fig. 1(a) shows the variation of the number of the participants who hold positive altitude and that of the participants who hold negative altitude. It indicates that the number of the participants with positive altitude decreases and then increases; on the contrary, the number of the participants with negative altitude increases and then decreases, and finally reaches to zero. So, all of the participants become the ones who have positive altitude. Fig. 1(b) presents the change of altitude for each participant. It demonstrates that the altitude of each participant, whether it is positive or negative, decreases firstly and then increases, and stops the variation at $\sigma = \pm 1$. And that an opinion consensus is formed. In this process, a part of the plus altitudes are dragged to be minus, and some even reach to $-1$. Where does the drag effect originate, then? The answer may lie in the structure of the environment: 446 members with positive altitude are slightly lesser than 454 members with negative altitude. The altitude inclination of the whole environment formed by 899 participants for any one is predetermined at the beginning. That is, no one can be immune from dragging by "negative" altitude of the environment. However, even if the altitudes of the large part of the participants successively reach to $-1$, at which the consensus of the opposite opinion likely arise, their values still turn to $+1$.

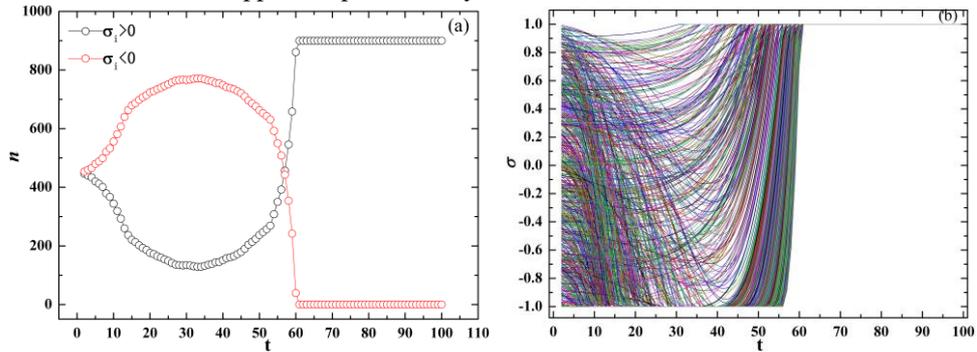

Figure 1 for the self-dominated driving case, (a) variation of number of the participants with the opposite altitudes and (b) variation of the altitude of each participant change with time.

Now, a further question may be asked: what causes the negative altitude to go back to the consensus point $+1$? The answer, of course, is the imbalance between the self-driving force and the environment-driving force. To understand this turn, we must reveal what happened at the turning point, t=36. Fig. 2 depicts the state that is characterized by the cognitive-level. Each of the participants is denoted by a point that locates on $x-\sigma$ space. Fig. 2(a) presents the initial state, the points randomly distribute in the plane. In order to pursue the change of the cognitive-level, some points, which locate at some special positions, are marked by different symbols with different color. To make clear the situation, Fig. 2(b) shows the situation at the 10-th time step so as to compare with that at 36-th time step. One may note that the violet-star points, which initially locate near to coordinate point (0,0) and are plus of values pierce "zero border" and move towards the lower-left corner; the gray-triangle points initially locate near the lower-right corner also move left decreasing their altitudes.

These change imply that the participants with lower cognitive-level and weaker altitude are apt to change their opinion due to the influence by circumstances. Meanwhile, the dark-green square points, which initially locate near to coordinate point (0, 14) in plus altitude region and those near to the top-right corner move slightly left. It indicates that the participants who with higher cognitive-level will stick to their colors even in the case that they have weaker altitude. Off course, the participants with higher cognitive-level and minus altitude will stick to their altitude (the opposite opinion). This situation is similar to that discussed for dark-green square points for the same reason. Generally, the altitude values of all the participants decrease for the "negative" altitude of the whole environment. And the negative altitudes are strengthened while the positive altitudes are weakened even some of them converted into negative ones.

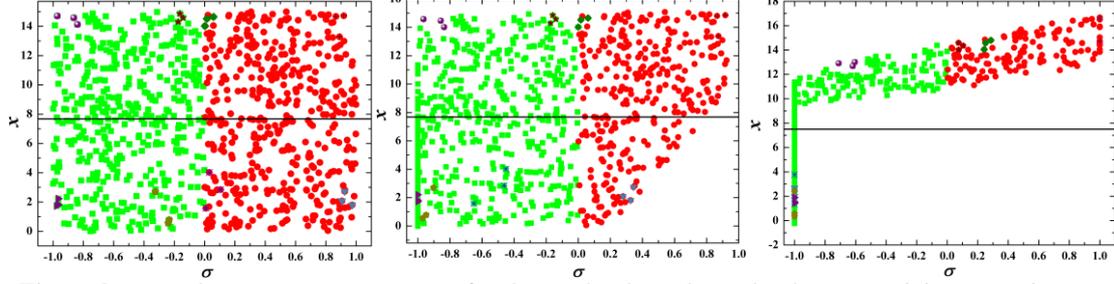

Figure 2 states shown on $x-\sigma$ space for the randomly and evenly chosen participants at time step (a) $t=1$, (b) $t=10$ and (c) $t=36$.

The trend continues until the 36-th time step. In this process, the altitude of the participants who have higher cognitive-level and stronger positive altitude firstly have rebounded after hitting rock bottom and reach to $+1$ since such strong judgment ability denoted by high cognitive-level strengthening such unbending attitude. In the final analysis this behavior, redounding after hitting rock bottom, can be only attributed to the self-driving force dominating the opinion evolution. Fig. 1(b) shows that some participants with negative altitude also have undergone the "rebounding after hitting rock bottom", which in fact is the forerunner of the altitude reversal. When it is at the 36-th time step, the first altitude reversal comes true, and is followed one by one as time evolves. And finally the opinion consensus is fulfilled at $\sigma=+1$. These discussions are also supported by the details presented by Fig. 2(c). Where the majority of the participants have high cognitive-level and the cognitive-level of the participant with positive altitude in general is higher than that of the participant with negative altitude. So the environment transforms into a "positive altitude" one. And it proves naturally that the self-driving force induce the "rebounding after hitting rock bottom" and lead to the altitude reversals.

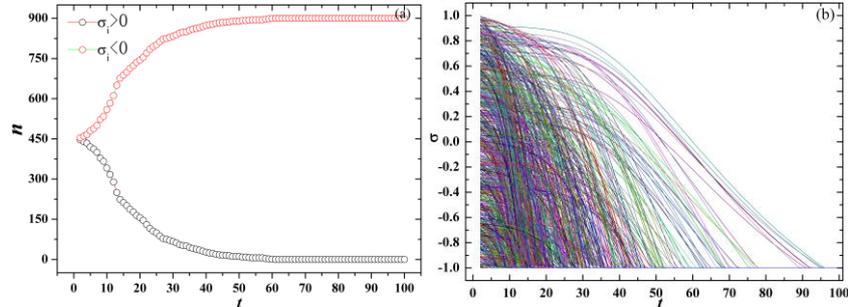

Figure 3 for the environment-dominated case, (a) variation of number of the participants with the opposite altitudes and (b) variation of the altitude of each participant change with time.

As far as the environment-dominated situation is concerned, the values of coefficients are chosen as $A=0.5\times10^{-3}, B=9.5\times10^{-3}$, and the results are presented in Fig. 3 and Fig. 4. Fig. 3(a) shows that the number of the participants with positive altitude decreases to zero, while the number of the participants with negative altitude increases. And finally the participants have fully become the ones with negative altitude. Fig. 3(b) presents that the altitudes of all participants monotonously decrease to $-1$, early or late. The details of the figure demonstrate that some participants initially have nearly same cognitive-levels exhibit different characteristics. Some have reduced their altitudes to $-1$ rapidly, and some others have reduced slowly. These differences are similar to that appeared in the preceding time steps in Fig. 1(b), and may result in the complexity of opinion evolution. One may ask: Why? Firstly, it is almost certain that the monotonicity of the altitudes varying downwards with time can be attributed to the fact that the opinion evolution is mainly dominated by the environment-driving force. The environment-driving force here is similar to the situations discussed in the first place in the self-dominated case And that environment is of "negative" altitude. The consensus of opinion can be naturally achieved at $\sigma=-1$. Secondly, the complexity of opinion evolution may be caused by fluctuations of the environment-driving force due to the fact that on phase space $x-\sigma$ some initially neighboring participants move away from each other. The typical representative is the violet-star points in Fig. 4(b) which separate after piercing the "zero border". Similar movements are denoted by the black-five-star points and dark-green-square points shown in Figs. 4 (b) and (c).

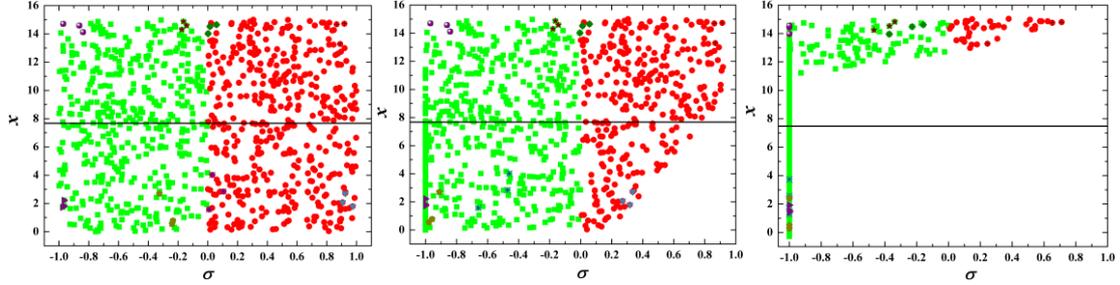

Figure 4 states shown on $x-\sigma$ space for the randomly and evenly chosen participants at time step (a) t=1 (it is the same as Fig. 2(a), and is re-shown here so as to compare with others), (b) t=10 and (c)t=36.

One may also notice that the participants who have low cognitive-level and weak altitude show a natural frangibility, that is, they are too apt to change their altitude. However, for the participants who have high cognitive-level and weak altitude, it not easy to change their altitude, for instance, the participants denoted by the black-five-star points and dark-green-square points. So, one can draw a conclusion that the cognitive-level plays a fundamental role in the opinion evolution process.

It may be interesting that if one separates n-t curve of negative altitude from that of positive altitude. The curve may be compared to a message propagation process in which the attention is excited, culminates, and declines at last to form a peak. The intensity of attention may be denoted by the number of participants. The profile of the peak serves as an impressive reminder of the message propagating on the internet, for instance, the comments induced by the event that a female driver was beaten in a traffic conflict in Chengdu city of China occurred on May 3, 2015. The comment number (actually represents the participant number) formed a pulse in the shape of a peak. It is shown by Fig. 5 and can be understood based on the similarity on growth and decline in the two processes.

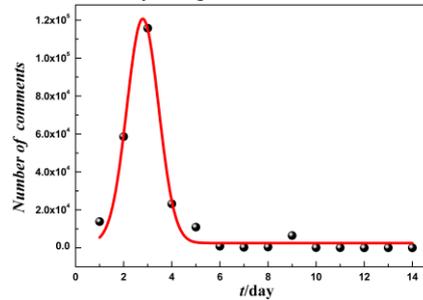

Figure 5 the comment number varies with time on the event that a female driver was beaten. The detail is presented in text.

## 3.2 Scaling properties of opinion formation

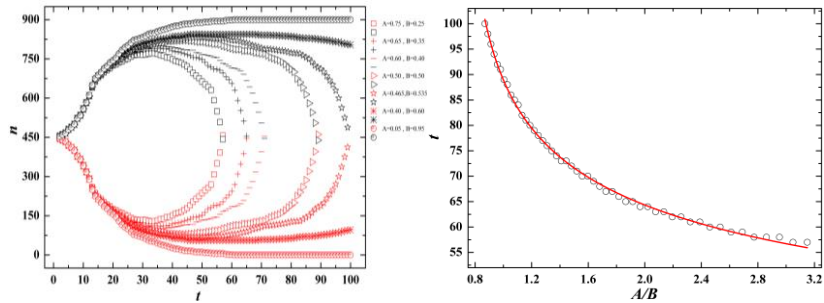

Fig. 6(a) the numbers of the participants with positive altitude and negative altitude change in different combination of A and B, (b) the dependence of $t_e$ on $A/B$, where the red curve is the exponential fitting.

As discussed above, the directions of opinion evolution are determined by the relative strength of the self-driving force and environment-driving force---the former is greater than the latter, or the opposite is the case. The first case covers a wide range including any combination of $A$ and $B$ as $A>B$, and also including the combination confined by $A/B<0.80$ as $A<B$. The dynamical detail of the opinion evolution is sensitive to the change of the combination. Fig. 6(a) shows the variation of the number of the participants with positive altitude and negative altitude in different combinations of $A$ and $B$. In order to highlight the turning points of increase-decrease respectively for the positive altitude and the negative altitude, we only present the sections before the time step, $t_e$, at which the numbers of the participants with the opposite altitudes are equal. It is interesting that time $t_e$ is correlated with ratio $A/B$ in a scaling manner. The scaling characteristic is demonstrated by Fig. 6(b), which indicates that $t_e$ depends on $A/B$ in an exponent way, i.e.,

$$t \propto e^{-\beta A/B}, \tag{9}$$

Where $\beta$ is a constant, and equal to $0.68$. This behavior implies that there often exists some simple regularity behind such a complex process, and also means that the stronger the influence from the self-diving force than that from the environment-driving force, the faster the consensus of opinion can be formed.

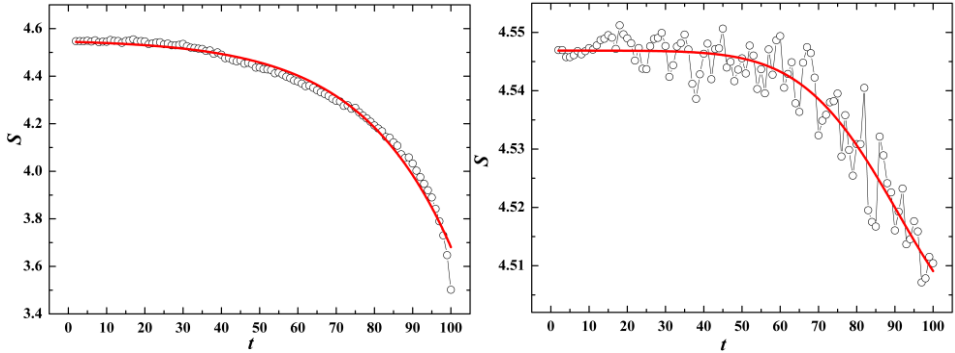

Figure 7 information entropy varies with time: (a) the typical case of self-driving force dominates the evolution of opinion; (b) the typical case of environment-driving force dominates the evolution of opinion.

From another point of view, the evolution of the altitudes of the participants to form a consensus of opinion is actually a ordering process, which can be described by variation of the so-called information entropy. The entropy can be defined as

$$S = -\sum_{l=1}^{m} p(l,t)\ln p(l,t) \tag{10}$$

Where $p(l,t)$ denotes the probability that the value of a cognitive-level is within the $l$-th interval. The intervals can be obtained via dividing the value range of the cognitive-level, $(0,15)$, into $m$ intervals uniformly. If one can find that there are $N(l,t)$ participants (among a total of $N$ participants) whose cognitive-level is within the $l$-th interval, then the probability $p(l,t)$ can be simply defined by the ratio of $N(l,t)$ to $N$, i.e.,

$$p(l,t) = \frac{N(l,t)}{N} \quad (11)$$

The results corresponding to the first case and the second case are depicted by Figs. 7(a) and (b), respectively. The fittings suggest that the entropy for the two process follow a same way decaying. That is, we have

$$S = S_0 + ae^{-\frac{(t-t_0)^2}{2\sigma^2}} \quad (12)$$

Where the parameters for Fig. 7(a) are $S_0 = 4.5, a = -4.2, \sigma = 1.1 \times 10^2, t_0 = 6.1 \times 10^2$ and for Fig. 7(b) are $S_0 = 4.5, a = -4.6 \times 10^{-2}, \sigma = 24.6, t_0 = 1.2 \times 10^2$. The fitting relation indicates that in the two processes the decaying of entropy means the system formed by the participants going into an order state. Of course, there is a little difference between the two processes lying in the details of the decaying way. The entropy in the first process decays smoothly with time, while the entropy in the second process decays in an oscillating mode. The cognitive-levels of the participants in general vary widely, which therefore can be regarded as being random. So, the oscillation may be attributed to the different contribution of each participant to the environment-driving force. This detailed difference between the entropy in the two processes may be the direct embodiment of the difference in the dynamics of opinion.

## 4 Conclusions and Discussions

In this article, we establish a model, a kinetic equation in a Newton-like form, to simulate dynamics of a public opinion. In the equation, the cognitive-level is introduced as the generalized displace; the resultant force is the summation of the self-driving force and the environment-driving force including the influences of each of the others act on the particular participant. The driving force is conventionally defined by the gradient of the so-called generalized potential, which is determined by the quantity of information that is related to one's cognitive-level and the altitude to the event, as well as the altitude collision between the particular and the one from the environment. According to the relative strength of the self-driving force and environment-driving force, the evolution of opinion can be classified into two types, self-dominated and environment-dominated.

This investigation revealed some regularity dominating the evolution of a public opinion. For the self-dominated driving case, the typical characteristics is the number of the participants with positive altitude increases after decreasing and that of the participant with negative altitude decreases after increasing. The preceding decreasing of the positive participants is mainly due to the negative altitude of the whole environment that is composed of more participants with negative altitude and lesser participants with positive altitude. For this reason, the participants initially with low cognitive-level and weak altitude will have their altitudes changed into negative ones for the lack of judgment. However, the altitude can rebound after hitting rock bottom because the participants who have positive altitude and high cognitive-level change the trend of altitude due to the fact that the strong self-driving force can get the jump on the environment-driving force. Therefore, the consensus of the opinion is fulfilled at last. For the environment-dominated case, the altitudes of all participants overwhelmingly evolve into the consensus to form an opposite opinion. It is no doubt a necessary consequence of the environment-dominated driving that each of the participants has the altitude of the environment.

The simulation reveals some regularity behind the dynamics of opinion. In the environment-dominated case, the time at which the number of the participants with positive altitude is equal to that of the participants with negative altitude scales with the ratio of the strength coefficients A/B in an exponent way. And the information entropy decays with time following the way that can be fitted by an exponent function of $(t-t_0)^2$. All the discussions may help us to understand the formation of a public opinion. Of course, more of essentially understandings may be expected.

## Acknowledgments

This study is supported by National Natural Science Foundation of China under Grant No. 11265011 and Ningxia Natural Science Foundation under Grant No. NZ12161.